\documentclass[12pt,a4paper]{article}
\usepackage[pdfstartview=FitH,colorlinks=true,linkcolor=black,anchorcolor=black,citecolor=blue,urlcolor=black]{hyperref}
\usepackage[english]{babel}
\usepackage[top=2.5cm,right=2.7cm,left=2.7cm,bottom=2.7cm]{geometry}
\usepackage{amsmath,amssymb,titling,authblk}
\usepackage{slashed}
\usepackage{amscd}
\usepackage{afterpage}
\usepackage{cite}
\usepackage{latexsym}
\usepackage{amsthm}
\usepackage{braket}
\usepackage{graphicx}
\usepackage{caption}
\usepackage{subcaption}
\usepackage{hyperref}

\allowdisplaybreaks[4]
\numberwithin{equation}{section}

\usepackage{color}  

\definecolor{verde}{cmyk}{.83,.21,1,.08}

\newcommand{\be}{\begin{equation}}
\newcommand{\ee}{\end{equation}}
\newcommand{\bea}{\begin{eqnarray}}
\newcommand{\eea}{\end{eqnarray}}

\begin{document}

\setlength{\droptitle}{-6pc}
\pretitle{\begin{flushright}\small
ICCUB-15-018
\end{flushright}\vspace*{2pc}%
\begin{center}\LARGE}
\posttitle{\par\end{center}}

\title{Green's Functions \\for Translation Invariant Star Products\vspace{5pt}}

\renewcommand\Affilfont{\itshape}
\setlength{\affilsep}{1.5em}
\renewcommand\Authands{ and }

\author[1,2,3]{Fedele Lizzi}
\author[2]{Manolo Rivera}
\author[1,2]{Patrizia Vitale}
\affil[1]{INFN, Sezione di Napoli\vspace{5pt}}
\affil[2]{Dipartimento di Fisica, Universit\`{a} di Napoli {\sl Federico II}\vspace{5pt}}
\affil[3]{Departament de Estructura i Constituents de la Mat\`eria, Institut \mbox{de Ci\'encies del Cosmos, Universitat de Barcelona}}

\date{}

\maketitle

\begin{abstract}\noindent
We calculate the Green functions for a scalar field theory with quartic interactions for which the fields are multiplied with a generic translation invariant star product. Our analysis involves both noncommutative products, for which there is the canonical commutation relation among coordinates, and nonlocal commutative products.  We give explicit expressions for the one-loop corrections to the two and four point functions. We find that the phenomenon of ultraviolet/infrared mixing is always a consequence of the presence of noncommuting variables. The commutative part of the product does not have the mixing.
\\
\\
\it{email: fedele.lizzi@na.infn.it, riveralam@gmail.com, patrizia.vitale@na.infn.it}

\end{abstract}

\medskip

%%% ======================================================================

\section{Introduction}

In this paper we study some aspects of field theory on noncommutative spaces, where the noncommutativity in encoded in the product of functions (and in particular fields) which are multiplied with a possibly noncommutative $\star$-product~\cite{generalstar1, generalstar2}. 
We will consider a  $\phi^{\star4}$ field theory, and as product we consider generalizations of the well known (and well studied) Gr\"onewold-Moyal~\cite{gron,moyal} product. For a general review of field theory with such product see for example~\cite{Szabo}. In particular we will make a complete study for the case of Translational Invariant Star Products (TISP).  These products have been introduced in~\cite{gallucciolizzivitalemixing} and studied also in~\cite{TanasaVitale, ArdalanSadoogi, Varshovi1, LizziVitalereg, Varshovi2}.  They all reproduce the standard commutation relation
	\begin{equation}
	[x^i,x^j]_{\star} =x^i\star x^j-x^j\star x^i= i \theta^{ij}. \label{commutator}
	\end{equation}
with $\theta^{ij}$ a constant. In this class of products, the case of $\theta=0$ is included. In such a case the product is commutative, but this does not mean that it is the usual pointwise one, it may be generalized to a nonlocal one.

Our study will center on the calculation of the Green function of the theory, and in particular we will calculate explicitly the one loop corrections of  the two and four points functions, i.e.\ the corrections to the propagator and vertex. In these two cases the Moyal star product manifests itself with a difference between planar and nonplanar diagrams, together with the loss of invariance under permutation of the external legs, and with the presence of Infrared/Ultraviolet mixing~\cite{IR/UVmixing}. 

The paper is organized as follows. In Section \ref{sectiongtisp} we review the derivation of translation invariant star products. In Section \ref{phi4} we derive the propagator and vertex for a real scalar field theory with quartic interaction and in Section \ref{green} we compute two and four points Green's functions up to one loop. We conclude with a short discussion section.
\section{The general translation invariant star products} \label{sectiongtisp}

In this section, following~\cite{gallucciolizzivitalemixing} we introduce the most general translation invariant star products in $d$ dimensions  that satisfy the associativity condition. We will not discuss the domain of definition of the various products, it is understood that some form or regularity is required, but we will not go into details. Written as an integral a generic product reads 
 	\begin{equation}
	(f \star g)(x)= \int \frac{d^dp}{(2\pi)^d} \frac{d^dq}{(2\pi)^d} \frac{d^dr}{(2\pi)^d} \tilde f(q) \tilde g(r) K(p,q,r) e^{-ip \cdot x}
	\end{equation}
where $\tilde f,\tilde g$  denote the Fourier transforms of $f$ and  $g$ respectively, and  $K$ is in general a distribution. Note that this includes commutative products, and in particular the ordinary pointwise product, which corresponds to $K(p,q,r) =(2\pi)^d \delta^{(d)}(r-p+q)$. 

The requirement of translation invariance imposes the condition  
 	\begin{equation}
	\mathcal T_a (f \star g)= \mathcal T_a(f) \star \mathcal T_a(g)
	\end{equation}
where $\mathcal T_a (f)(x) = f(x+a)$ is the translation by any vector $a \in \mathbb R^d$. The left hand side is given by 
 	\begin{equation}
	\mathcal T_a (f \star g)= \int \frac{d^dp}{(2\pi)^d} \frac{d^dq}{(2\pi)^d} \frac{d^dr}{(2\pi)^d} \tilde f(q) \tilde g(r) K(p,q,r) e^{-ip \cdot (x+a)}
	\end{equation}
while the right hand side is 
 	\begin{align}
	\mathcal T_a(f) \star \mathcal T_a(g) &= \int \frac{d^dp}{(2\pi)^d} \frac{d^dq}{(2\pi)^d} \frac{d^dr}{(2\pi)^d} \widetilde { \mathcal T_a (f)}(q) \widetilde { \mathcal T_a (g)}(r) K(p,q,r) e^{-ip \cdot x} \nonumber\\
	 &= \int \frac{d^dp}{(2\pi)^d} \frac{d^dq}{(2\pi)^d} \frac{d^dr}{(2\pi)^d} e^{ia \cdot q} \tilde f(q) e^{ia \cdot r} \tilde g(r) K(p,q,r) e^{-ip \cdot x}.
	\end{align}
Thus the product is translation invariant if these two expressions are equal for any vector $a$, which means
 	\begin{equation}
	  K(p,q,r) =(2\pi)^d e^{\alpha (p,q)} \delta^{(d)}(r-p+q)
	\end{equation}
where $\alpha$ is a generic (possibly complex) function. So the general translation invariant star product is given by
 	\begin{equation}
	(f \star g)(x)= \int \frac{d^dp}{(2\pi)^d} \frac{d^dq}{(2\pi)^d} \tilde f(q) \tilde g(p-q) e^{\alpha (p,q)} e^{-ip \cdot x}. \label{36}
	\end{equation}
In particular, the well known Moyal product corresponds to the following  function $\alpha$
 	\begin{equation}
	   \alpha_M (p,q) = \frac{i }{2}\theta^{ij} p_i q_j,
	\end{equation}
so, from now on we will indicate with  $ \alpha_M$ the  part of $\alpha$ which reproduces the Moyal case.

There are three constraints that the function $\alpha$ has to satisfy. 

\begin{description} 
	\item[(i)]  Associativity: 
 	\begin{equation}
	 ((f \star g) \star h)(x) =  (f \star (g \star h))(x) .
	\end{equation}
	\item[(ii)] Multiplication by a constant function is the ordinary scalar multiplication:
 	\begin{equation}
	(f \star 1)(x) =\int \frac{d^dp}{(2\pi)^d} \tilde f(q) e^{\alpha (p,p)} e^{-ip \cdot x}=f \label{73}
	\end{equation}
and

 	\begin{equation}
	(1 \star f)(x) =\int \frac{d^dp}{(2\pi)^d} \tilde f(q) e^{\alpha (p,0)} e^{-ip \cdot x}=f \label{74}
	\end{equation}
	\item[(iii)] The algebra of functions with the star product is a $*$-algebra. That is, there must be a map which satisfies the following conditions:
 	\begin{align}
	(f^*)^* &= f \\
	(\lambda f+ \mu g)^* &= \bar \lambda f^* + \bar \mu g^* \\
	( f \star g)^* &=   g^*  \star f^*
	\end{align}
where $\lambda,\mu \in \mathbb C$, and the bar denotes complex conjugation. In this case the involution $*$ is given by complex conjugation.
\end{description}
Note that condition (iii) is a natural condition if we want to represent the functions as operators once we study the field theory of noncommutative spaces corresponding to these star products.  

It is not difficult to check that these three conditions give the following constraints on $\alpha$
\begin{description} 
\item[(i)] Associativity: 
 	\begin{equation}
	 \alpha (q,t)+\alpha (p,q)=\alpha (p-t,q-t)+\alpha (p,t) \label{associativity}
	\end{equation}
\item[(ii)] Multiplication by a constant function:
 	\begin{align}
		\alpha (p,p) &=0 \nonumber\\
		\alpha (p,0)&=0 	
	\end{align}
\item[(iii)] Involution:
 	\begin{equation}
		 \alpha(p,q)^*=\alpha(-p,q-p)
 	\end{equation}
\end{description} 
It has been shown in~\cite{ArdalanSadoogi} that the most general function that satisfies these conditions is of the form
 	\begin{equation}
	 \alpha(p,q) = \eta(q) - \eta(p) + \eta(p-q) +  i\omega(p,q) \label{37}
	\end{equation}
where $\omega(p,q) = \frac{1 }{2}\theta^{ij} p_i q_j$, and  $\eta(p)=\eta_1 (p) + i\eta_2 (p)$ with $\eta_1$ an arbitrary real and even function  such that $\eta_1(0)=0$, and $\eta_2$ a real odd function without linear term, i.e.
 	\begin{equation}
	 \eta_2(p) = \sum_{n=1}^{\infty} \sum_{l=0}^{2n+1} C_{l,2n+1-l} \, p_1^l p_2^{2n+1-l}.
	\end{equation}
The $\eta$ function corresponding to the Moyal product is given by $\eta(p) = 0$.

It is easy to show that the commutativity condition, $f \star g = g \star f$,  requires
 	\begin{equation}
	\alpha(p,q) = \alpha (p,p-q)
	\end{equation}
but note that from \eqref{37} we have
 	\begin{equation}
	\alpha(p,q) =  \eta(q) - \eta(p) + \eta(p-q) +  i\omega(p,q)
	\end{equation}
and
 	\begin{equation}
	\alpha(p,p-q) =  \eta(q) - \eta(p) + \eta(p-q) +  i\omega(p,p-q).
	\end{equation}
Therefore, just $\omega$ contributes to the non-commutativity of the product. But note also that $i\omega(p,q)=\alpha_M(p,q) $, so it is just the Moyal part of $\alpha$ what contributes to the non-commutativity of the product. Heuristically, any noncommutative TISP is the Moyal product ``plus" some (non-local) commutative product, while any commutative TISP is of the form $\alpha(p,p-q) =  \eta(q) - \eta(p) + \eta(p-q)$. This also implies that any noncommutative TISP satisfies the commutator $[{x}^i,{x}^j]_{\star} = i \theta^{ij}$.   

%----------------------------------------------------------------------  Quantum Field Theory with a general translation invariant star product--------------------------------------
\section{Quantum Field Theory with a general translation invariant star product}\label{phi4}
We now discuss the noncommutative field theory with a general translation invariant star product. For definiteness and simplicity we  consider the $(2+1)$-dimensional space-time where the two spacial coordinates satisfy the commutator \eqref{commutator}, and the time coordinate commutes with the two spacial coordinates. The higher dimensional cases can be easily built. Consider the action:
	\begin{equation}
	S = S_0 - S_{\text{int}}  \label {7}
	\end{equation}
where $S_0$ is the free Klein-Gordon action given by 
	\begin{equation}
	S_0 = \int d^3x \frac{1}{2}(\partial_\mu \phi  \star \partial^\mu \phi - m^2 \phi  \star \phi) \label{S0}
	\end{equation}
while the interacting action becomes
	\begin{equation}
	S_{\text{int}} =  \frac{g}{4!} \int d^3x \; \phi  \star \phi  \star \phi  \star \phi. \label{8}
	\end{equation}
To compute the equations of motion we will need the following property
 	\begin{eqnarray}
	\int d^3 x \, f \star g &=& \int d^3 x \, \frac{d^3p}{(2\pi)^3}\frac{d^3q}{(2\pi)^3} \tilde f(q) \tilde g(p-q) e^{\alpha (p,q)} e^{-ip \cdot x} \nonumber \\
	&=&  \int \frac{d^3q}{(2\pi)^3} \tilde f(q) \tilde g(-q) e^{2\eta_1(q)}, \label{40}
	\end{eqnarray}
which in particular implies the necessary trace property for the integral:
  	\begin{equation}
	\int d^3 x \, f  \star g= \int d^3 x \; g  \star f. \label{traceprop}
	\end{equation}  
Doing a small variation of the field $\phi \to \phi + \delta \phi $, the variation of the action is 
	\begin{equation}
	\delta S_0 = \int d^3x (\partial_\mu \delta \phi  \star \partial^\mu \phi - m^2 \delta \phi  \star \phi).
	\end{equation}
where we used property \eqref{traceprop}. Integrating by parts and neglecting the boundary terms this is
	\begin{equation}
	\delta S_0 = - \int d^3x  \delta \phi  \star ( \Box + m^2) \phi . \label{38}
	\end{equation}
Using equation \eqref{40} we have
 	\begin{equation}
	\delta S_0 = - \int \frac{d^3q}{(2\pi)^3} \widetilde {\delta \phi}(q)  (-q^2 + m^2)\tilde \phi(-q) e^{2\eta_1(q)} 
	\end{equation}
but this must vanish for any variation of the field $\delta \phi$, so the equation of motion in momentum space is
 	\begin{equation}
	 e^{2\eta_1(q)}  (q^2 - m^2)\tilde \phi(q) = 0 \label{39}
	\end{equation}
%%
%this is equivalent to the ordinary equation
% %
%	\begin{equation}
%	  ( q^2 - m^2)  \tilde{\phi} (q)= 0
%	\end{equation}
%%
Since the exponential is never zero, at  the classical level, the noncommutative free field theory given by the action~\eqref{S0} with any translation invariant star product, is the same as the commutative one.
We now proceed to the computation of the propagator and the vertex. The propagator can be easily found from  \eqref{39} to be 
 	\begin{equation}
	 \tilde G_0 (p) = \frac{ e^{-2\eta_1(q)} }{ (p^2 - m^2)}. \label{propagator}
	\end{equation}
The vertex can be computed using equations \eqref{36} and \eqref{40}
 	\begin{align}
	 S_{\text{int}} =& \frac{g}{4!} \int d^3x \prod_{a=1}^4 \frac{d^3k_a}{(2\pi)^3}\,
	 %\frac{d^3k_1}{(2\pi)^3}\frac{d^3k_2}{(2\pi)^3}\frac{d^3k_3}{(2\pi)^3}\frac{d^3k_4}{(2\pi)^3} 
	 \tilde \phi(k_2) \tilde \phi(k_1-k_2) \tilde 	 \phi(k_4) \tilde \phi(k_3-k_4) \nonumber \\ &  \hspace{16em}   e^{\alpha(k_1,k_2)}e^{\alpha(k_3,k_4)}e^{k_1 \cdot x}\star e^{k_3 \cdot x} \nonumber \\
	  = &\frac{g}{4!} \int \prod_{a=1}^4 \frac{d^3k_a}{(2\pi)^3}\,\tilde \phi(k_2) \tilde \phi(k_1-k_2) \tilde \phi(k_4) \tilde \phi(k_3-k_4) \nonumber \\ & \qquad e^{\alpha(k_1,k_2)}e^{\alpha(k_3,k_4)}  \int \frac{d^3k}{(2\pi)^3}e^{\alpha(0,k)}(2\pi)^3 \delta^{(3)}(k_1 - k) (2\pi)^3 \delta^{(3)}(k_3 + k)  \nonumber \\
	  = &\frac{g}{4!}(2\pi)^3 \int \prod_{a=1}^4 \frac{d^3k_a}{(2\pi)^3}\,\tilde \phi(k_2) \tilde \phi(k_1-k_2) \tilde \phi(k_4) \tilde \phi(k_3-k_4) \nonumber \\ & \hspace{12em}  e^{\alpha(k_1,k_2) + \alpha(k_3,k_4) + \alpha(0,k_1)} \delta^{(3)}(k_1 + k_3) 
	\end{align}
which can be written as 
 	\begin{eqnarray*}
	 S_{\text{int}}  &=& \frac{g}{4!}(2\pi)^3 \int \prod_{a=1}^4 \frac{d^3k_a}{(2\pi)^3}\,\tilde \phi(k_1) \tilde \phi(k_2) \tilde \phi(k_3) \tilde \phi(k_4) \nonumber \\&& e^{\alpha(k_1+k_2,k_2) + \alpha(k_3+k_4,k_4) + \alpha(0,k_1+k_2)} \delta^{(3)}(k_1 +k_2+ k_3+k_4) 
	\end{eqnarray*}
So the vertex is given by
 	\begin{equation}
	V_{\star} = V e^{\alpha(k_1+k_2,k_2) + \alpha(k_3+k_4,k_4) + \alpha(0,k_1+k_2)} \label{41}
	\end{equation}
where
	\begin{equation}
	 V = -i \frac{g(2\pi)^3}{4!}  \delta^{(3)}(k_1+k_2+k_3+k_4) \label{101}
	\end{equation}
 is the ordinary vertex. 
 
Recalling that  $\omega(p,q) = \frac{1 }{2}\theta^{ij} p_i q_j$ we have $- \eta(k_3 + k_4)+ \eta(-k_1-k_2)=0$ and $ \omega(k_2,k_1)  +\omega(k_4,k_3) =- \sum_{a<b} \omega(k_a,k_b)$, using equation \eqref{37}, 
% as
%%
% 	\begin{align}
%	V_{\star} = V e^{\eta(k_1) - \eta(k_1 + k_2) + \eta(k_2) +  i\omega(k_1+k_2,k_1) } e^{ \eta(k_3) - \eta(k_3 + k_4) + \eta(k_4) +  i\omega(k_3+k_4,k_3) } \\  e^{ \eta(k_1+k_2) - \eta(0) + \eta(-k_1-k_2) +  i\omega(0,k_1+k_2) }
%	\end{align}
% 	\begin{equation}
%	= V e^{\eta(k_1) + \eta(k_2)+ \eta(k_3) + \eta(k_4) +  i\omega(k_2,k_1)  - \eta(k_3 + k_4)  +  i\omega(k_4,k_3) + \eta(-k_1-k_2) }
%	\end{equation}
%%
%But using 
and conservation of momentum in the vertex $-k_1-k_2 = k_3 + k_4$, the vertex \eqref{41} can be written,
 	\begin{equation}
	V_{\star}= V e^{\sum_a \eta(k_a) -   i\frac{\theta}{2} \sum_{a<b} k_a \wedge k_b} \label{42}
	\end{equation}
where $p \wedge q=\varepsilon^{ij}p_iq_j=p_1 q_2 - p_2 q_1$ and  $\theta^{ij}=\theta \varepsilon^{ij}$. The vertex is not invariant under arbitrary permutations of the lines entering it. The Moyal factor $ i\frac{\theta}{2} \sum_{a<b} k_a \wedge k_b$ is invariant under cyclic permutations
%\footnote{Here by cyclic permutation we mean $\sigma \in S_n$ of the form $\sigma(a) = (a + k)mod\, n$} 
but it may not be invariant under non-cyclic permutations. On the other hand, the factor $\sum_a \eta(k_a)$ is invariant under any permutation. This means that the symmetries of the vertices are totally determined by the Moyal part.

%-------------------------------------------------------------------------------------- Green's functions for a general translation invariant star product --------------------------------
\section{Green's functions for a general translation invariant star product}\label{green}

Having obtained the vertex and  the propagator,  we can compute the two-point and four-point Green functions. We will be very detailed  in this section, showing how to obtain all symmetry factors and showing explicitly all calculations.

At zeroth order we have the same diagrams as in conventional quantum field theory (QFT) except that the propagators are given by~\eqref{propagator}. At first order, some things are different because we already have a vertex (which as we saw, is not invariant under a generic permutation), so we need to keep track of the order in which the lines enter the vertex. This is why some diagrams that are equivalent in ordinary QFT, are in general different in Noncommutative Quantum Field Theory (NCQFT). We will proceed as in the usual QFT case (see for example\cite{peskin}) 
 	\begin{align}
	(2\pi)^3 \delta^{(3)}(\sum_{i=1}^N k_i) G^{(N)}(k_1,...,k_N) =(2\pi)^{3N/2} \text{exp} \left( -iS_{\text{int}} \left(\frac{\delta}{i\,\delta j(-q)} \right) \right)  \times & \nonumber \\  \frac{\delta^N}{i\,\delta j(-k_1) \dots i\,\delta j(-k_N)} \; \text{exp} \left(- \frac{1}{2} \int d^3 k \; j(k) G (k) j(-k) \right)& \label{32}
	\end{align}
where 
 	\begin{equation}
	S_{\text{int}} \left(\frac{\delta}{i\,\delta j(-q)} \right) = i \int \left[ \prod_{i=1}^4 \frac{d^3 q_i}{(2\pi)^{3/2}} \frac{\delta}{\delta j(-q_i)} \right]  V_{\star}.
	\end{equation}
The computation of the Green functions is done in the same way as in QFT, but we have to keep track of the order of the lines in each vertex, due to the phase  $ i\frac{\theta}{2} \sum_{a<b} k_a \wedge k_b$ in $V_{\star}$. 

Note that it is possible to consider quantum field theories with a translation invariant star product without Moyal part. But we saw that just the Moyal part contributes to the non-commutativity of the product. This means that a translation invariant product without Moyal part would give rise to a commutative (possibly non-local) quantum field theory, in which case we would have the same diagrams as in conventional QFT. Here we are mainly interested in the noncommutative products, which give rise to the commutator \eqref{commutator}. See however \cite{LizziVitalereg, AAGS13} where a non-local commutative product has been suggested as an alternative tool for regularization of UV divergences.

In the following sections we will compute the  connected component of the Green's function $G_c$, which is the relevant part for  scattering amplitudes.

%------------------------------------------- two-point greens function----------------------------------------------------------------
\subsection{Two-point Green's function} \label{2pointgf}

In order to see how all this works, we will do the computation of the connected two-point Green's function $\tilde G_c^{(2)}$ in detail. From equation \eqref{32} we can see that $\tilde G_c^{(2)}(p)$ (where $p=k_1=-k_2$) at leading order is
 	\begin{equation}
	\tilde G_0^{(2)}(p) = \tilde G_0 (p) = \frac{e^{-2\eta_1(q)} }{ (p^2 - m^2)}.
	\end{equation}
where the superscript refers to the order of expansion. It is just the propagator, because there is no vertex. At one loop, things get more complicated than in the commutative case because of the vertex.

\begin{figure}[htb]
\hspace*{1cm}
	\begin{subfigure}{0.45\textwidth}
		\includegraphics[scale=0.45]{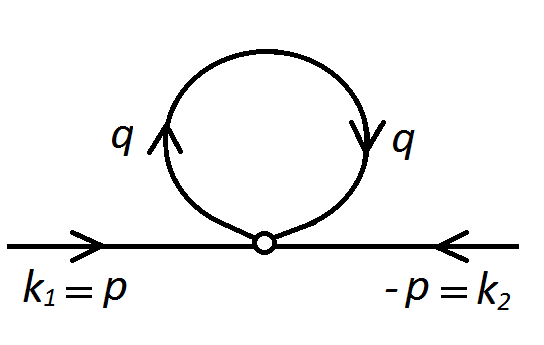}
		\caption{} \label{fig1a}
	\end{subfigure}
	\begin{subfigure}{0.45\textwidth}
		\includegraphics[scale=0.45]{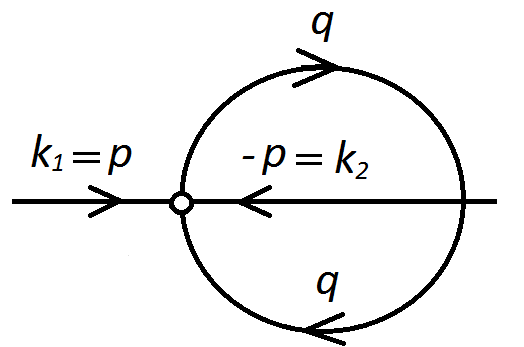}
		\caption{} \label{fig1b}
	\end{subfigure}
	\caption{\sl Two-point Green's function one loop diagrams} \label{fig1}
\end{figure}

\begin{figure}[htb]
	\begin{subfigure}{0.3\textwidth}
		\hspace*{1cm}
		\includegraphics[scale=0.4]{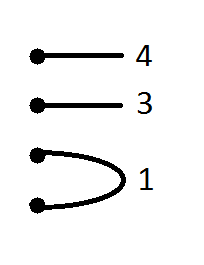}
		\caption{} \label{fig2a}
	\end{subfigure}
	\begin{subfigure}{0.3\textwidth}
		\hspace*{1cm}
		\includegraphics[scale=0.4]{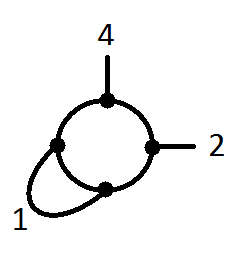}
		\caption{} \label{fig2b}
	\end{subfigure}
	\begin{subfigure}{0.3\textwidth}
		\hspace*{1cm}
		\includegraphics[scale=0.4]{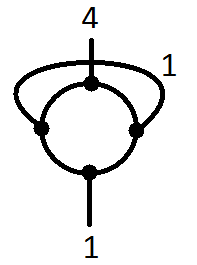}
		\caption{} \label{fig2c}
	\end{subfigure}

	\caption{\sl Two-point Green's function vertex. Figure (a) corresponds to the vertex in QFT, figures (b) and (c) correspond to the two inequivalent vertices in NCQFT. The numbers represent the number of ways of attaching the corresponding line to the vertex} \label{fig2}
\end{figure}

 In QFT we would just have the diagram shown in figure \ref{fig1a}
with a symmetry factor $12/4!$. The usual way to compute this symmetry factor is by thinking of the vertex as made by four points, and then count the number of ways to attach the four lines to the vertex, as it is shown in figure \ref{fig2a}.
In NCQFT we do the same but we just count the number of ways of attaching the lines in such a way that the vertex doesn't change. Due to the invariance under cyclic permutations, just the order of the lines is relevant, so we can think of the four points as being on a ring, like in figures \ref{fig2b} and \ref{fig2c}.  Saying that the vertex is invariant under cyclic permutations is like saying that rotating the ring doesn't change anything. Let us use this to compute  $\tilde G_c^{(2)}$. Using the fact that $k_1=-k_2$ it is easy to see that there are just two inequivalent vertices, shown in figures \ref{fig2b} and \ref{fig2c}, and their corresponding diagrams are shown in figures \ref{fig1a} and \ref{fig1b} respectively. The symmetry factor of \ref{fig2b} is $8/4!$ because there are four ways of attaching the first external line, two ways of attaching the second external line (to the right or to the left of the first one) and the internal loop joins the two remaining points. On the other hand, the symmetry factor of \ref{fig2c} is $4/4!$ because there are four ways for the first external line, the second one must be in the opposite point and the internal loop joins the two remaining points. As a check one can verify that the sum of the two symmetry factors is $12/4!$, which is the symmetry factor in QFT.    

The integral corresponding to a given diagram is found in the same way as in ordinary QFT (as can be seen from equation \eqref{32}), i.e.\ the Feynman rules are the same except that the symmetry factors are computed in the way we explained, and the vertex and propagator are given by equations \eqref{42} and \eqref{propagator}. For the diagram \ref{fig1a}, the vertex is given by\footnote{To compute the phase of the vertex we choose any line (due to cyclic invariance) and then move counterclockwise} 
	\begin{align}
	 V_{\star} &= V e^{\eta(p)+\eta(-p)+\eta(q)+\eta(-q) -i\frac{\theta}{2} (-\boldsymbol q \wedge \boldsymbol q +\boldsymbol q\wedge \boldsymbol p -\boldsymbol q\wedge \boldsymbol p -\boldsymbol q\wedge \boldsymbol p +\boldsymbol q\wedge \boldsymbol p -\boldsymbol p \wedge \boldsymbol p)}\nonumber\\
	&=V e^{2(\eta_1(p)+\eta_1(q)) }. 
	\end{align}
The integral corresponding to the diagram \ref{fig1a} is then given by
	\begin{align}
	 \tilde{G}_{\ref{fig1a}}^{(2)}(p)&=\frac{-ig}{3} \int \frac{d^3 q}{(2\pi)^3}\, \frac{ e^{-( 4\eta_1(p)+ 2\eta_1(q))}  e^{2(\eta_1(p)+\eta_1(q)) } }{ (p^2 - m^2)^2 (q^2 - m^2)} \nonumber \\
	&= \frac{-ig}{3} \int \frac{d^3 q}{(2\pi)^3}\, \frac{  e^{- 2\eta_1(p)} }{ (p^2 - m^2)^2 (q^2 - m^2)} \label{2order2pointa}
	\end{align}
Doing the same for the diagram  \ref{fig1b} we find
	\begin{equation}
	  \tilde{G}_{\ref{fig1b}}^{(2)}(p)= \frac{-ig}{6} \int \frac{d^3 q}{(2\pi)^3}\, \frac{  e^{- 2\eta_1(p) - i\theta \boldsymbol p\wedge \boldsymbol q} }{ (p^2 - m^2)^2 (q^2 - m^2)} \label{2order2pointb}
	\end{equation}
The connected two-point Green's function up to one loop order is the propagator plus the sum of these two integrals with their corresponding symmetry factors
 	\begin{equation}
	\tilde G_{c; \,1}^{(2)}(p) =  \frac{e^{-2\eta_1(q)} }{ (p^2 - m^2)}- \frac{ig}{6} \int \frac{d^3 q}{(2\pi)^3}\, \frac{( 2\, +  e^{- \theta i \boldsymbol p\wedge \boldsymbol q}) e^{- 2\eta_1(p)}  }{  (p^2 - m^2)^2 (q^2 - m^2)} .
	\end{equation}
%
%Note that if we substitute the $\eta$ function corresponding to the ordinary product ($\eta = 0$), we get the ordinary expression 
% 	\begin{equation}
%	 \tilde{G}_{c; \,1}^{(2)}(p)=\frac{ 1}{ p^2 - m^2} - \frac{ig}{2} \int \frac{d^3 q}{(2\pi)^3}\, \frac{1 }{  (p^2 - m^2)^2 (q^2 - m^2)} .
%	\end{equation}

%------------------------------------------------ Four point green's function --------------------------------------------------------------
\subsection{Four-point Green's function} \label{4pointgf}

\begin{figure}[htb]
	\centering
	\includegraphics[scale=0.4]{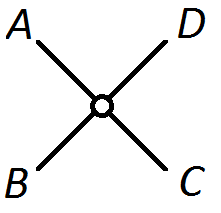}
	\caption{\sl Four-point Green's function diagrams with one vertex. There are 6 inequivalent diagrams of this type, corresponding to: $ ABCD \mapsto  k_1 k_2 k_3 k_4$ (1),  $\mapsto k_1 k_3 k_2 k_4$ (2),  $\mapsto k_1 k_4 k_2 k_3$ (3), $ \mapsto k_1 k_2 k_4 k_3$ (4), $\mapsto k_1 k_3 k_4 k_2$ (5) ,  $ \mapsto k_1 k_4 k_3 k_2$(6).} \label{fig3}
\end{figure}

\begin{figure}[htb]
	\begin{subfigure}{0.5\textwidth}
\hspace*{2.2cm}
		\includegraphics[scale=0.4]{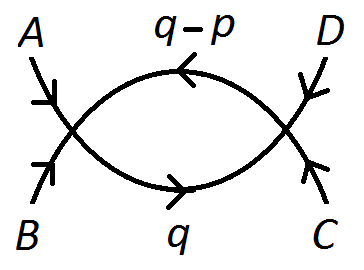}
		\caption{} \label{fig4a}
	\end{subfigure}
	\begin{subfigure}{0.5\textwidth}
\hspace*{2.5cm}
		\includegraphics[scale=0.38]{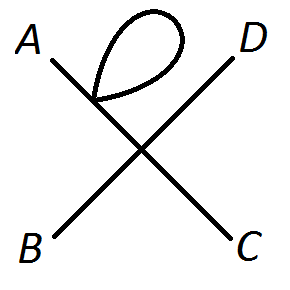}
		\caption{} \label{fig4b}
	\end{subfigure}
	\caption{\sl Four-point Green's function diagrams in QFT at one loop order, where $p=k_A+k_B$.} \label{fig4}
\end{figure}

\begin{figure}[htb]
%\renewcommand*\thesubfigure{\arabic{subfigure}}
%\makeatletter
%\renewcommand{\p@subfigure}{\thefigure-}
%\makeatother
	\begin{subfigure}{0.3\textwidth}
	%	\hspace*{1cm}
		\includegraphics[scale=0.35]{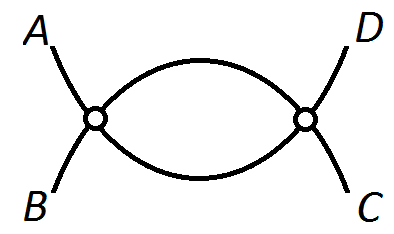}
		\caption{} \label{fig50}
	\end{subfigure}
	\begin{subfigure}{0.3\textwidth}
	%	\hspace*{1cm}
		\includegraphics[scale=0.35]{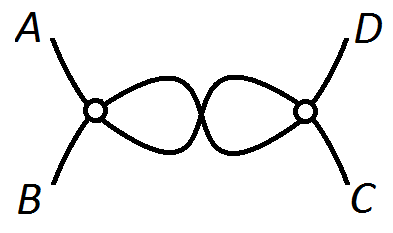}
		\caption{} \label{fig5a}
	\end{subfigure}
	\begin{subfigure}{0.3\textwidth}
%		\hspace*{1cm}
		\includegraphics[scale=0.35]{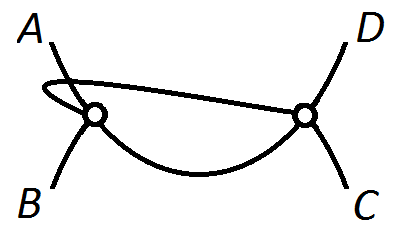}
		\caption{} \label{fig5b}
	\end{subfigure}
	\begin{subfigure}{0.3\textwidth}
%		\hspace*{1cm}
		\includegraphics[scale=0.35]{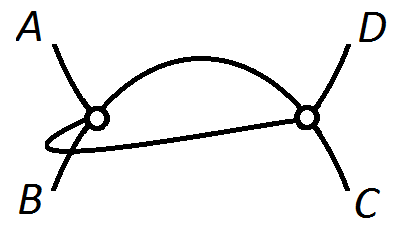}
		\caption{} \label{fig5c}
	\end{subfigure}
	\begin{subfigure}{0.3\textwidth}
%		\hspace*{2cm}
		\includegraphics[scale=0.35]{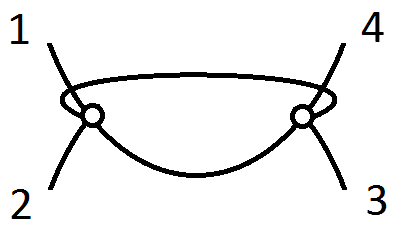}
		\caption{} \label{fig5d}
	\end{subfigure}
	\begin{subfigure}{0.3\textwidth}
%		\hspace*{2cm}
		\includegraphics[scale=0.35]{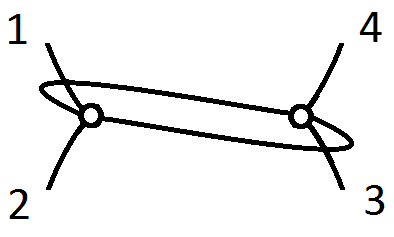}
		\caption{} \label{fig5e}
	\end{subfigure}
	\caption{\sl Four-point Green's function diagrams in NCQFT at one loop order.} \label{fig5}
\end{figure}

We now compute the four-point Green's function. At first order we have the ordinary diagram of the form shown in figure \ref{fig3},
 but in NCQFT, the six different permutations of this diagram are not equivalent, so we have to consider them separately. Each of these diagrams has a symmetry factor of $4/4!$ because there are four ways to attach the first external line to the vertex, and then the other three are completely determined. Recalling that   $\sum_a \eta(k_a)$ is invariant under any permutation, we can easily see that we have
 	\begin{align}
	 \tilde{G}_{\ref{fig3}\text{-}j}^{(4)}(k_1,k_2,k_3,k_4)  =  -i \frac{g}{6} \frac{ e^{-\sum_a \bar \eta(k_a)-\frac{i\theta}{2} E_j}}{\prod_a ( k_a^2 - m^2)}   \label{1order4point}
	\end{align}
where the subindex $j \in \{ 1,\dots 6\}$ corresponds to each of the 6 diagrams, $E_j$ is the Moyal part of the vertex 
\begin{equation}
E_j=k_A\wedge k_B + k_A\wedge k_C + k_A\wedge k_D + k_B\wedge k_C + k_B\wedge k_D + k_C\wedge k_D
\end{equation}
of the corresponding diagram with the assignment of $\{ABCD\}$ as in the figure caption, and $\bar \eta$ is the complex conjugate of $\eta$. 
At one loop order, the only connected diagrams we would have in QFT are shown in figure \ref{fig4}.
 Let us first look at diagrams \ref{fig4a}.
In QFT there are three inequivalent diagrams of this type, corresponding to $ABCD \mapsto k_1k_2k_3k_4$, $\mapsto k_1k_3k_2k_4$ and $\mapsto k_1k_4k_3k_2$. Each of these diagrams has a symmetry factor of $1/2$. In NCQFT, for each of these three diagrams there are 18 different diagrams, which are of 6 different typologies and are shown in figure \ref{fig5}.
 To each of these diagrams we associate a phase $F_{n,m}(A,B,C,D)$, where the first intex $n\in\{1,2,3\}$ corresponds to the three types of commutative diagrams in figure \ref{fig4a} and the second index $m\in\{1,...,6\}$ corresponds to the 6 types of noncommutative diagrams in figure \ref{fig5}. Let us look at the 18 noncommutative diagrams that correspond to \ref{fig4a} with $ABCD \mapsto k_1k_2k_3k_4$. First there are 4 planar diagrams like the one in figure \ref{fig50}, with $ABCD \mapsto k_1k_2k_3k_4$,  $\mapsto k_2k_1k_3k_4$, $\mapsto k_1k_2k_4k_3$ and  $\mapsto k_2k_1k_4k_3$. To these diagrams we associate a phase 
 \be
 F_{n,1}(A,B,C,D)=k_A \wedge k_B + k_C \wedge k_D
 \ee
There other 14 diagrams are of 5 different typologies which are shown in Figure~\ref{fig5}. There are four diagrams of the form \ref{fig5a} where  $ABCD \mapsto k_1k_2k_3k_4$, $\mapsto k_2k_1k_3k_4$,  $\mapsto k_1k_2k_4k_3$ or  $\mapsto k_2k_1k_4k_3$. Their corresponding phase is 
 \be
 F_{n,2}(A,B,C,D)=k_A \wedge k_B + k_C \wedge k_D+ 2(k_A+k_B) \wedge q
 \ee
 There are four diagrams of the form \ref{fig5b} with  $ABCD \mapsto k_1k_2k_3k_4$, $\mapsto k_1k_2k_4k_3$,  $\mapsto k_4k_3k_1k_2$ or  $\mapsto k_4k_3k_2k_1$ and phase 
 \be
 F_{n,3}(A,B,C,D)=k_A \wedge k_B + k_C \wedge k_D- 2k_A \wedge q
 \ee 
There are four diagrams of the form \ref{fig5c} with  $ABCD \mapsto k_1k_2k_3k_4$, $\mapsto k_1k_2k_4k_3$,  $\mapsto k_4k_3k_1k_2$ or  $\mapsto k_4k_3k_2k_1$  and phase 
\be
F_{n,4}(A,B,C,D)=k_A \wedge k_B + k_C \wedge k_D+ 2k_B \wedge q
\ee
Finally, there is one diagram of the form \ref{fig5d} and one of the form \ref{fig5e} whose phases are respectively given by 
\be 
F_{n,5}\,=  -k_1 \wedge k_2 - k_3 \wedge k_4 
\ee
 and \be
F_{n,6}=  k_1 \wedge k_2  + k_3 \wedge k_4 +2(k_A+k_B) \wedge q\
\ee

The symmetry factor of each of these 18 diagrams is equal to $(4 \cdot 4 )/(4! \cdot 4!)=1/(2\cdot 18)$. To see this, note that in this case the only symmetries of the diagrams are the cyclic permutations of the two vertices, so the first external line of each vertex has four options, and then the positions of the other three lines of each vertex are completely determined (each non-cyclic permutation gives rise to an inequivalent diagram). The sum of the symmetry factors of the diagrams in figure \ref{fig5} is equal to the symmetry factor of diagram \ref{fig4a}, as expected.  

Using again the fact that $\sum_a \eta(k_a)$ is invariant under any permutation, we can see that the non-Moyal part of the two vertices is given by 
 	\begin{equation}
	 \eta (k_1)+ \eta (k_2)+ \eta (q)+ \eta(p_n- q)+ \eta (k_3)+ \eta (k_4)+ \eta (-q)+ \eta (q- p_n)
	\end{equation}
 	\begin{equation}
	= \sum_a \eta (k_a)+ 2\eta_1 (q)+ 2\eta_1(p_n- q) 
	\end{equation}
where the index $n\in\{1,2,3\}$ is a label for the three diagrams \ref{fig4a}, and the corresponding values of $p_n$ are: $p_1=k_1+k_2$, $p_2=k_1+k_3$ and $p_3=k_1+k_4$. The last two factors of this expression cancel with the two exponentials coming from the propagators of the two internal lines. With this, the corresponding integral of each of the diagrams in figure \ref{fig5} are found to be
 	\begin{equation}
	 \tilde{G}_{\ref{fig5}\text{-}nm}^{(4)}(k_A,k_B,k_C,k_D) =\frac{-g^2}{36} \int \frac{d^3 q}{(2\pi)^3} \frac{ e^{-\sum_a \bar\eta(k_a)-i\frac{\theta}{2} F_{nm}(A,B,C,D)} }{(q^2-m^2)((p_n-q)^2-m^2)\prod_a ( k_a^2 - m^2)}
	\end{equation}
where the first subindex $n$, as we said, is a label for the three commutative diagrams \ref{fig4a}, while the second subindex $m$ is a label for the 6 types of noncommutative diagrams in figure \ref{fig5}. More explicitly, $\tilde{G}_{\ref{fig5}\text{-}1m}$ for $m\in \{ 1,2,\dots ,6 \}$ corresponds to the 6 diagrams that reduce to% 
\ref{fig4a} with  $ABCD \mapsto k_1k_2k_3k_4$, in the commutative limit, and equivalently $\tilde{G}_{\ref{fig5}\text{-}2m}$ and $\tilde{G}_{\ref{fig5}\text{-}3m}$ correspond to the diagrams that reduce to \ref{fig4a} with $ABCD\mapsto k_1k_3k_2k_4$ and $ABCD\mapsto k_1k_4k_3k_2$ respectively.

\begin{figure}[htb]
	\begin{subfigure}{0.5\textwidth}
\hspace*{2.2cm}
		\includegraphics[scale=0.35]{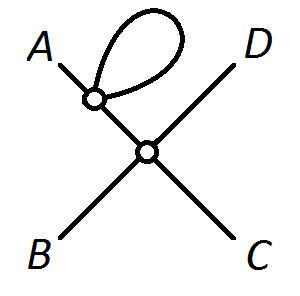}
		\caption{} \label{fig7a}
	\end{subfigure}
	\begin{subfigure}{0.5\textwidth}
\hspace*{2.5cm}
		\includegraphics[scale=0.35]{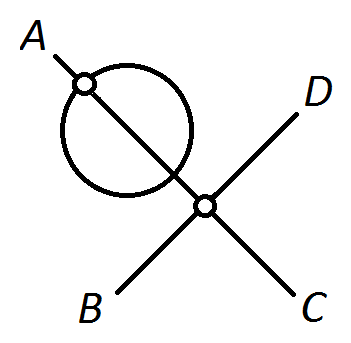}
		\caption{} \label{fig7b}
	\end{subfigure}
	\caption{\sl Four-point Green's function diagrams in NCQFT at one loop order. There are 6 of type (a) and 6 of type (b).} \label{fig7}
\end{figure}

Finally, let us look at diagrams \ref{fig4b}. In QFT there are four inequivalent diagrams corresponding to $A \mapsto k_i$, $i\in\{1,2,3,4\}$. In NCQFT, for each of these diagrams, there are 12 different diagrams shown in figure \ref{fig7}.
 The diagrams in figure ~\ref{fig7a} have a symmetry factor of $(4\cdot4\cdot2)/(4!\cdot4!)=1/18$ and their corresponding integrals are given by  
 	\begin{equation}
	 \tilde{G}_{\ref{fig7}a\text{-}ij}^{(4)}(k_A,k_B,k_C,k_D) = \frac{(-ig)^2}{18} \int \frac{d^3 q}{(2\pi)^3} \frac{ e^{-\sum_a \bar \eta(k_a)-i\frac{\theta}{2} E_j }  \;  }{(q^2-m^2)(k_i^2-m^2)\prod_a ( k_{a}^2 - m^2)} 
	\end{equation} 
where the $E_j$ is the same we had in equation (\ref{1order4point}), and again the subindex $i\in\{1,\dots,4 \}$  is a label for the four commutative diagrams coresponding to  $A \mapsto k_i$, while the index $j\in\{1,\dots ,6 \}$ is a label for the six noncommutative diagrams (of the form shown in figure \ref{fig7}) that reduce to the corresponding commutative diagram (i.e. $ \tilde{G}_{\ref{fig7}a\text{-}ij}$ for  $j\in\{1,\dots ,6 \}$ correspond to the 6 noncommutative diagrams that reduce to \ref{fig4b} with $A \mapsto k_i$, in the commutative limit).

Finally, the diagrams in figure \ref{fig7b} have a symmetry factor of $(4\cdot4)/(4!\cdot4!)=1/36$ and their corresponding integrals are given by  
 	\begin{equation}
	 \tilde{G}_{\ref{fig7}b\text{-}ij}^{(4)}(k_A,k_B,k_C,k_D) = \frac{(-ig)^2}{36} \int \frac{d^3 q}{(2\pi)^3} \frac{ e^{-i \theta \boldsymbol k_i \wedge \boldsymbol q}\, e^{-\sum_a \bar \eta(k_a)-i\frac{\theta}{2} E_j }  \;  }{(q^2-m^2)(k_i^2-m^2)\prod_a ( k_{a}^2 - m^2)} 
	\end{equation}
where  $E_j$ is the same as before, as well as the explanation of the subindices. We can finally write the connected four-point Green's function up to one loop order. For this we have to sum the contributions of all the diagrams:  
 	\begin{align}
	 \tilde{G}_{c;2}^{(4)}(k_1,k_2,k_3,&k_4)   =\sum_{j=1}^6 \tilde{G}_{\ref{fig3}\text{-}j}^{(4)} +  \sum_{n=1}^3\sum_{\{m\}}\tilde{G}_{\ref{fig5}\text{-}nm}^{(4)} +\sum_{i=1}^4\sum_{j=1}^6\tilde{G}_{\ref{fig7}a\text{-}ij}^{(4)} +  \sum_{i=1}^4\sum_{j=1}^6\tilde{G}_{\ref{fig7}b\text{-}ij}^{(4)}\nonumber\\ 
	=-& \sum_j  \frac{ig}{6} \frac{ e^{-\sum_a \bar \eta(k_a)-\frac{i\theta}{2} E_j}}{\prod_a ( k_a^2 - m^2)} \nonumber\\
	-& \sum_{n\{m\}}\frac{g^2}{36} \int \frac{d^3 q}{(2\pi)^3} \frac{ e^{-\sum_a \bar\eta(k_a)-i\frac{\theta}{2} F_{nm}(A,B,C,D)} }{(q^2-m^2)((p_n-q)^2-m^2)\prod_a ( k_a^2 - m^2)} \nonumber \\
	-& \sum_{ij}\frac{g^2}{36} \int \frac{d^3 q}{(2\pi)^3} \frac{\left( 2+ e^{-i \theta \boldsymbol k_i \wedge \boldsymbol q} \right) \, e^{-\sum_a \bar \eta(k_a)-i\frac{\theta}{2} E_j }  \;  }{(q^2-m^2)(k_i^2-m^2)\prod_a ( k_{a}^2 - m^2)}
	\end{align}
	\label{finalresult}
where the notation $\sum_{\{m\}}$ means that $m$ runs from 1 to 6, but we have to take into account the four inequivalent diagrams for the cases $m=1,2,3$ and $4$. 
It is not difficult to see that if we set $\eta=0$ we get back the usual expression for the connected four-point Green's function in QFT. 

Note that in all cases the limit $\theta\to 0$, i.e.\ the case of a commutative non local product, there is only a correction of the kind $e^{-\sum_a \bar\eta(k_a)}$. This term could be used as a regulator, along the lines of~\cite{LizziVitalereg, AAGS13}.

\section{Discussion}

We have calculated the Green's functions at one loop for a generic translation invariant product in a $\phi^{*4}$ theory. We considered both one particle irreducible and reducible diagrams. We have shown that the corrections factorize in the one  that we would get for the Gr\"onewold-Moyal product, and another term which has to do with the commutative part. The latter corrections do not show the infrared/ultraviolet mixing, and show no difference between planar and nonplanar diagrams. The mixing is due to the oscillating phase $e^{i\theta k_i\wedge q}$ in the internal loop interactions appearing in~\eqref{finalresult} and in the equations leading to it. Being an oscillating term it softens the ultraviolet divergence, but is also responsible for a novel infrared divergence.

Our result confirms earlier results that the mixing phenomenon is a direct consequence of the presence of a Moyal part of the product, and is heuristically related to a ``space uncertainty'' of the particular form~\cite{DFR}. Our results also generalize those obtained  for the Wick-Voros product~\cite{gallucciolizzivitaletwist} (see also~\cite{Balvoros1, Balvoros2}).

Since the translation invariant products generalize  the Gr\"onewold-Moyal and Wick-Voros products properly, it is natural to consider if a generalization of the twisted symmetries that they enjoy~\cite{Wess, chaichian, book,  aschierilizzivitaletwistall, book} can be generalized as well. This is currently under investigation and will appear in a future publication~\cite{LizziRiveraVitalesymm}.

\subsubsection*{Acknowledgments}
Part of this work was performed as part of an internship at the University of Napoli done by M.R.\ as part of his master degree at the University of Marseille/Aix. We thank both institutions for this opportunity.
This article is based upon work from COST Action MP1405 QSPACE, supported by COST (European Cooperation in Science and Technology).
F.L.\ is partially supported by CUR Generalitat de Catalunya under projects FPA2013-46570 and 2014~SGR~104, MDM-2014-0369 of ICCUB (Unidad de Excelencia 'Maria de Maeztu') .  F.L.\ and P.V.\ are partially supported by UniNA and Compagnia di San Paolo under the grant Programma STAR 2013.

\end{document}